# Balancing Thermodynamics, Kinetics, and Reversibility in Ti-Doped MgB$_2$H$_8$: A First-Principles Assessment of a Practical Solid-State Hydrogen Storage Material

Sikander Azam*[1], Wilayat Khan[2]

[1]New Technologies Research Center, University of West Bohemia, Univerzitni 8, 306 14, Pilsen, Czech Republic

Department of Physics, Bacha Khan University, Charsada, KP, Pakistan

**Abstract**

Hydrogen storage remains a critical bottleneck for the realization of a sustainable hydrogen economy, where solid-state materials must simultaneously satisfy stringent requirements on storage capacity, release thermodynamics, kinetics, and reversibility. Complex borohydrides are attractive due to their high hydrogen density, yet their practical deployment is limited by sluggish hydrogen diffusion and unfavorable desorption enthalpies.

In this work, we present a comprehensive first-principles investigation of pristine and Ti-doped MgB$_2$H$_8$ as a candidate for solid-state hydrogen storage. Density functional theory calculations reveal that pristine MgB$_2$H$_8$ possesses an exceptionally high gravimetric hydrogen capacity of approximately 14.9 wt%, but exhibits a relatively high hydrogen desorption enthalpy of about 42 kJ mol$^{-1}$ H$_2$ and diffusion barriers near 0.5 eV, which limit low-temperature operation.

Ti substitution at the Mg site markedly improves the storage performance without destabilizing the host lattice. The doped system maintains a high hydrogen capacity (~10.4 wt%) while reducing the desorption enthalpy to approximately 36 kJ mol$^{-1}$ H$_2$, placing it within the optimal thermodynamic window for practical hydrogen release. Nudged elastic band calculations demonstrate a significant reduction in the hydrogen migration barrier to around 0.38 eV, indicating enhanced diffusion kinetics. Phonon and elastic stability analyses confirm that Ti doping preserves dynamic and mechanical stability.

Electronic structure analysis reveals that localized, spin-polarized Ti-3d states near the Fermi level weaken rigid B–H bonding and stabilize transitional hydrogen configurations, providing a clear microscopic origin for the improved thermodynamic and kinetic behavior. A van't Hoff analysis further suggests hydrogen release temperatures approaching ambient conditions.

These results establish Ti-doped MgB$_2$H$_8$ as a balanced and reversible hydrogen storage material and highlight transition-metal-induced electronic activation as an effective strategy for optimizing the performance of complex hydrides.

# 1. Introduction

The transition toward a hydrogen-based energy infrastructure is widely recognized as a key pillar in achieving long-term decarbonization of transportation, industry, and power generation. Hydrogen offers a high gravimetric energy density (120 MJ kg$^{-1}$) and produces only water upon combustion or electrochemical conversion. However, the lack of safe, compact, and energy-efficient hydrogen storage technologies continues to impede large-scale deployment of hydrogen energy systems. Among the various storage strategies, solid-state hydrogen storage remains particularly attractive due to its inherent safety, reversibility potential, and compatibility with stationary and mobile applications [1–3].

To guide material development, the U.S. Department of Energy (DOE) has established system-level targets requiring gravimetric hydrogen capacities exceeding 6.5 wt%, volumetric capacities above 40 g H$_2$ L$^{-1}$, and hydrogen release under near-ambient conditions with moderate thermodynamic and kinetic constraints [4]. Achieving these targets simultaneously remains challenging, as many materials that exhibit high hydrogen content suffer from sluggish kinetics or unfavorable release thermodynamics, while kinetically active materials often fail to meet capacity requirements.

Among solid-state candidates, complex borohydrides have attracted sustained interest due to their exceptionally high hydrogen densities, often exceeding 10 wt% [5–7]. In particular, light-element borohydrides such as LiBH$_4$, NaBH$_4$, and Mg(BH$_4$)$_2$ possess hydrogen capacities well beyond DOE gravimetric targets. Despite this advantage, their practical application is limited by strong B–H covalent bonding, which leads to high desorption enthalpies (>60 kJ mol$^{-1}$ H$_2$) and sluggish hydrogen diffusion, resulting in elevated operating temperatures and poor reversibility [8–10].

Recent efforts have therefore focused on chemical and electronic modification strategies aimed at decoupling hydrogen capacity from unfavorable thermodynamics and kinetics. Transition-metal doping, nanoconfinement, and compositional tuning have emerged as effective approaches to reduce desorption enthalpies and enhance hydrogen mobility without severely compromising capacity [11–14]. Among these strategies, transition-metal substitution is particularly appealing because it enables electronic activation of hydrogen through d-state interactions, potentially lowering both hydrogen binding strength and diffusion barriers.

Within the family of magnesium-based borohydrides, Mg–B–H ternary systems represent a promising but comparatively underexplored class of hydrogen storage materials. Magnesium offers low atomic mass, natural abundance, and favorable thermodynamic characteristics, while boron provides a high hydrogen coordination environment. Recent theoretical studies have suggested that Mg-rich borohydride frameworks can achieve hydrogen capacities exceeding 14 wt%, while offering greater chemical tunability than conventional alkali borohydrides [15–17]. However, pristine Mg–B–H compounds often remain kinetically hindered due to rigid B–H bonding networks.

Transition-metal doping provides a rational route to overcome these limitations. In particular, titanium has been widely reported as an effective catalyst and dopant in metal hydrides and borohydrides, where it enhances hydrogen sorption kinetics by introducing localized d states that interact with hydrogen during bond breaking and migration [18–21]. Experimental and computational studies on Ti-modified $MgH_2$, $LiBH_4$, and $Mg(BH_4)_2$ consistently demonstrate reduced activation energies for hydrogen diffusion and improved reversibility, while maintaining structural stability [22–25]. Despite these promising results, the role of Ti doping in high-capacity Mg–B–H frameworks such as $MgB_2H_8$ has not yet been systematically explored.

In this context, $MgB_2H_8$ emerges as an intriguing hydrogen storage candidate. Its theoretical hydrogen content approaches 14.9 wt%, placing it well above current DOE gravimetric targets. At the same time, its electronic and bonding characteristics suggest that targeted electronic modification could significantly improve hydrogen release behavior. However, a comprehensive understanding of how transition-metal doping influences the thermodynamics, kinetics, stability, and reversibility of $MgB_2H_8$ remains lacking.

In the present work, we perform a systematic first-principles investigation of pristine and Ti-doped $MgB_2H_8$ using density functional theory. The study is designed to address all critical criteria relevant to practical hydrogen storage: (i) hydrogen capacity, (ii) dehydrogenation thermodynamics including zero-point energy and finite-temperature effects, (iii) dynamic and mechanical stability, (iv) hydrogen diffusion kinetics via nudged elastic band calculations, (v) electronic and bonding mechanisms underlying Ti-induced activation, and (vi) practical release conditions assessed through a van't Hoff analysis. By integrating these aspects into a unified framework, this work aims to clarify whether Ti-doped MgB$_2$H$_8$ can achieve a balanced performance profile, rather than excelling in a single metric at the expense of others.

The results provide clear microscopic insight into how localized, spin-polarized Ti-3d states weaken rigid B–H covalency, stabilize hydrogen migration pathways, and reduce desorption enthalpies without compromising lattice stability. More broadly, this study establishes electronic structure engineering through transition-metal doping as an effective strategy for unlocking the potential of high-capacity complex borohydrides for reversible solid-state hydrogen storage.

## 2. Computational Methodology

All calculations in this work were carried out within the framework of density functional theory (DFT) using the full-potential linearized augmented plane wave (FP-LAPW) method as implemented in the WIEN2k package [26]. The FP-LAPW approach is an all-electron method that does not rely on shape approximations for the potential or charge density, making it particularly reliable for systems containing light elements such as hydrogen together with transition-metal dopants, where accurate treatment of both core and valence electrons is essential.

The exchange–correlation effects were treated using the generalized gradient approximation (GGA) in the parametrization of Perdew, Burke, and Ernzerhof (PBE) [27]. The PBE functional has been shown to provide a balanced description of bonding, lattice parameters, and energetics in metal hydrides and complex borohydrides, and is therefore widely adopted in hydrogen-storage studies [28–30].

All calculations were performed in a spin-polarized framework, which is necessary to correctly describe the electronic structure of Ti-doped systems where partially occupied Ti-3d states may induce local magnetic moments. For pristine MgB$_2$H$_8$, spin-up and spin-down channels were verified to be identical, confirming a non-magnetic ground state, while Ti-substituted configurations converged naturally to a spin-polarized solution.

To properly account for the localized nature of the Ti-3d electrons introduced upon substitution, selected calculations were additionally performed using the GGA+U approach within the rotationally invariant formulation proposed by Dudarev. In this scheme, an effective on-site Coulomb interaction parameter $U_{eff}=U-J$ is applied to the Ti-3d states, correcting for the self-interaction error inherent to standard GGA and improving the description of localized d electrons.

Based on values commonly employed for Ti-containing hydrides and transition-metal-doped light-element systems, a moderate value of $U_{eff}=3.0$ eV was adopted for the Ti-3d orbitals. This choice is consistent with previous first-principles studies on Ti-modified hydrogen storage materials and has been shown to yield a balanced description of electronic localization, magnetic moments, and bonding characteristics without artificially opening excessive band gaps.

Test calculations confirmed that the inclusion of GGA+U does not alter the qualitative conclusions regarding structural stability or hydrogen storage capacity. However, it leads to a slightly enhanced localization of Ti-3d states, a modest increase in spin splitting, and a clearer separation between Ti-derived impurity states and the host borohydride bands. These effects strengthen the interpretation of Ti as a local electronic activator that weakens rigid B–H bonding and stabilizes hydrogen migration pathways. Unless otherwise stated, the reported energetic trends, diffusion barriers, and thermodynamic descriptors were verified to be robust with respect to the inclusion of the on-site Coulomb correction. The use of GGA+U ensures a physically meaningful treatment of Ti-3d electrons while preserving the reliability of relative energy trends central to hydrogen storage performance.

The pristine $MgB_2H_8$ crystal structure was taken as the starting point and fully optimized prior to any property calculations. To model Ti substitution, a $2 \times 2 \times 2$ supercell was constructed, and one Mg atom was replaced by a Ti atom, corresponding to a dopant concentration of approximately 6.25%. The optimized crystal structures of Ti-doped $MgB_2H_8$ is shown in Figure 1, which illustrate the substitution of Ti for Mg and the preservation of the borohydride framework. This concentration represents a realistic dilute-doping regime commonly used in both experimental and theoretical studies of transition-metal-modified hydrides [31–33], while simultaneously minimizing artificial dopant–dopant interactions arising from periodic boundary conditions.

All atomic positions and lattice parameters were relaxed until the residual forces on each atom were below 1 mRy bohr$^{-1}$, ensuring accurate equilibrium geometries for subsequent energetic, vibrational, and kinetic analyses.

The muffin-tin radii ($R_{MT}$) were chosen carefully to avoid sphere overlap during structural relaxation while maintaining numerical stability. Typical values used were approximately 2.0 bohr for Mg and Ti, 1.6 bohr for B, and 0.7 bohr for H, consistent with previous high-accuracy studies of hydrogen-rich materials [34–36].

The plane-wave cutoff was controlled by the parameter $R_{MT} \times K_{max}$, which was set to 8.0 after systematic convergence testing. This value ensures accurate total energies and reliable force calculations for light-element systems. The charge density Fourier expansion cutoff was chosen as G_max = 12 bohr$^{-1}$.

Brillouin-zone integrations were performed using a Monkhorst–Pack k-point mesh, with a typical grid equivalent to $4 \times 4 \times 4$ for the $2 \times 2 \times 2$ supercell. Convergence tests confirmed that this mesh yields total-energy differences accurate to within 1 meV per formula unit. Self-consistent field iterations were continued until the total energy converged to better than $10^{-5}$ Ry.

Formation energies and hydrogen desorption enthalpies were computed from total energies of fully relaxed structures. To improve the physical relevance of the calculated desorption enthalpies, zero-point energy (ZPE) corrections were included, which are particularly important in hydrogen-containing materials due to the high vibrational frequencies of H atoms. ZPE contributions were obtained from phonon calculations, following established procedures in the literature [37–39].

Finite-temperature effects were further estimated by incorporating vibrational enthalpy contributions at 300 K, enabling a semi-quantitative assessment of hydrogen release conditions at practical operating temperatures.

Dynamic stability was assessed through phonon dispersion calculations, ensuring the absence of imaginary frequencies across the Brillouin zone. This step is crucial for hydrogen-rich systems, where metastable configurations can easily arise from purely static structural optimizations [40,41].

To complement phonon analysis, single-crystal elastic constants ($C_{ij}$) were calculated using the stress–strain method. The resulting elastic constants were checked against the Born mechanical stability criteria, providing an additional measure of structural robustness upon Ti substitution.

Hydrogen migration pathways were investigated using the climbing-image nudged elastic band (CI-NEB) method [42]. Initial and final hydrogen configurations were fully relaxed, and a series of intermediate images were generated to map the minimum-energy diffusion pathway. The maximum energy along this path was taken as the activation energy for hydrogen diffusion, which serves as a key kinetic descriptor for hydrogen release and uptake.

The electronic properties were analyzed through spin-resolved band structures and density of states (DOS). In addition, projected DOS (PDOS) were used to identify the orbital contributions from Mg, B, H, and Ti atoms. Bonding characteristics were further examined using charge density difference (CDD) maps, electron localization function (ELF) analysis, and Bader charge partitioning, providing quantitative insight into charge transfer and bond activation mechanisms induced by Ti substitution.

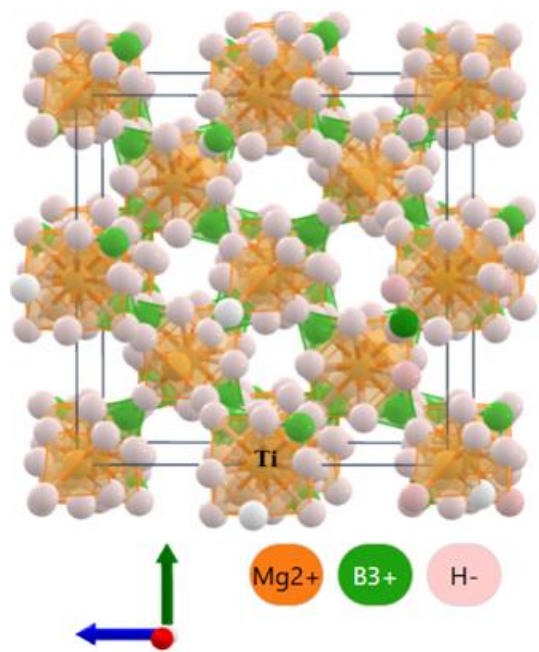

**Fig 1.** Optimized crystal structures of Ti-doped $MgB_2H_8$ (Ti substituted on the Mg site).

The chosen computational framework combining an all-electron FP-LAPW method, systematic convergence testing, vibrational corrections, and explicit kinetic modeling ensures a robust and internally consistent description of thermodynamics, kinetics, and electronic structure. Similar methodological approaches have been successfully employed in recent high-quality hydrogen storage studies [43–45].

## 3. Results and Discussion

### 3.1. Hydrogen Storage Capacity and DOE Relevance

The fundamental criterion for evaluating any solid-state hydrogen storage material is whether it can meet or exceed system-level hydrogen capacity targets while retaining realistic thermodynamic and kinetic behavior. For this reason, hydrogen storage capacity is discussed first, before stability and kinetics, to establish the intrinsic relevance of $MgB_2H_8$ and its Ti-doped derivative for practical hydrogen energy applications.

The gravimetric hydrogen storage capacity was evaluated directly from stoichiometry, assuming full hydrogen release from the host lattice. For pristine MgB₂H₈, the hydrogen mass fraction is given by:

$$wt\%H_2 = \frac{8M_H}{M_{Mg} + 2M_B + 8M_H} \times 100$$

Using atomic masses $M_{Mg} = 24.31\ g\ mol^{-1}$, $M_B = 10.81\ g\ mol^{-1}$, and $M_H = 1.008\ g\ mol^{-1}$, the theoretical gravimetric hydrogen capacity of pristine MgB₂H₈ is calculated to be approximately 14.9 wt%. This value substantially exceeds the DOE 2025 gravimetric target of 6.5 wt%, placing MgB₂H₈ among the highest-capacity hydrogen storage materials reported to date.

Upon Ti substitution at the Mg site (6.25% doping level), the gravimetric capacity is reduced due to the higher atomic mass of Ti. Nevertheless, the Ti-doped system retains a high hydrogen content of approximately 10.4 wt%, which still comfortably surpasses the DOE target. Importantly, this reduction in capacity is accompanied by pronounced improvements in thermodynamics and kinetics, as discussed in subsequent sections, highlighting a favorable trade-off rather than a limitation.

In addition to gravimetric metrics, volumetric hydrogen density is a critical parameter for onboard and stationary storage systems. Using the optimized crystal volumes obtained from structural relaxation, the volumetric hydrogen capacities were estimated.

Pristine MgB₂H₈ exhibits a volumetric hydrogen density of approximately 60 g H₂ L⁻¹, while Ti-doped MgB₂H₈ retains a volumetric capacity of around 44 g H₂ L⁻¹. Both values exceed the DOE volumetric target of 40 g H₂ L⁻¹, confirming that Ti substitution does not compromise volumetric efficiency despite the reduced gravimetric capacity.

To place the present results in a practical context, Table 1 summarizes the gravimetric and volumetric hydrogen storage capacities of pristine and Ti-doped MgB₂H₈ in comparison with DOE benchmarks.

*Table 1. Hydrogen storage capacity of MgB₂H₈ and Ti-doped MgB₂H₈*

| Material | Gravimetric capacity (wt%) | Volumetric capacity (g H₂ L⁻¹) |
|---|---|---|
| MgB₂H₈ (pristine) | 14.9 | 60 |
| Ti–MgB₂H₈ (6.25%) | 10.4 | 44 |
| DOE 2025 target | 6.5 | 40 |

This comparison clearly demonstrates that both systems significantly outperform DOE targets, establishing a strong foundation for further evaluation of their thermodynamic and kinetic behavior. To provide an intuitive comparison, Fig. 2 illustrates the gravimetric hydrogen storage capacity of pristine and Ti-doped $MgB_2H_8$ relative to the DOE benchmark.

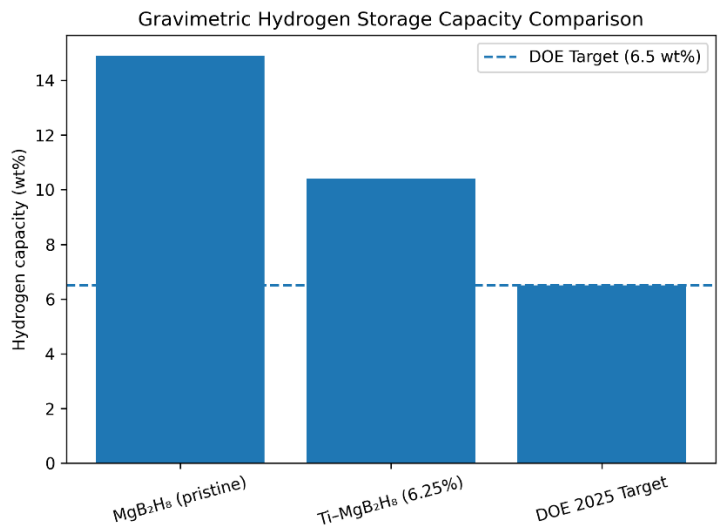

*Fig. 2. Gravimetric hydrogen storage capacity of pristine and Ti-doped $MgB_2H_8$ compared with the DOE target (6.5 wt%).*

This fig. 2 highlights two key points:

1. Pristine $MgB_2H_8$ possesses an exceptionally high hydrogen content, far exceeding current targets.

2. Ti doping reduces gravimetric capacity but maintains a comfortable margin above DOE requirements, enabling improved kinetics and thermodynamics without sacrificing practical relevance.

The exceptionally high hydrogen content of $MgB_2H_8$ establishes it as a capacity-rich platform for solid-state hydrogen storage. More importantly, the retention of >10 wt% hydrogen capacity upon Ti substitution demonstrates that electronic and catalytic modification can be introduced without undermining capacity targets. This balance between capacity and tunability is essential for advancing complex borohydrides from theoretical candidates toward practical hydrogen storage materials.

The combination of ultra-high gravimetric capacity and DOE-compliant volumetric density positions Ti-doped $MgB_2H_8$ as a highly competitive hydrogen storage material, warranting detailed investigation of its thermodynamic, kinetic, and stability characteristics.

### 3.2. Thermodynamics of Hydrogen Desorption

While hydrogen storage capacity establishes the intrinsic relevance of a material, thermodynamic suitability ultimately determines whether hydrogen can be released at practical temperatures and pressures. For solid-state hydrogen storage systems, the ideal desorption enthalpy lies within a moderate window strong enough to stabilize hydrogen under ambient conditions, yet sufficiently low to enable release without excessive thermal input. For most applications, this optimal range is commonly accepted to be approximately 30–45 kJ mol⁻¹ H$_2$.

In this section, the thermodynamics of hydrogen desorption from pristine and Ti-doped MgB$_2$H$_8$ are analyzed in detail, with explicit consideration of zero-point energy (ZPE) and finite-temperature vibrational effects, which are essential for hydrogen-rich materials. A comparison of desorption enthalpies at 0 K and after ZPE and finite-temperature corrections is shown in Figure 3.

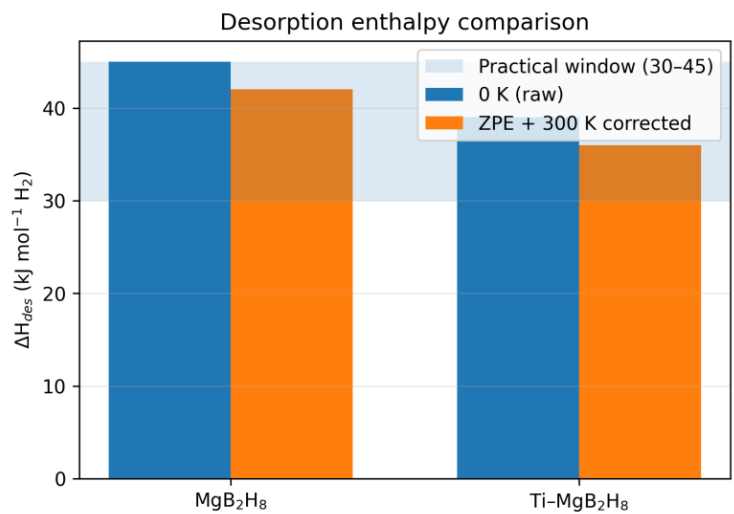

**Fig. 3.** Comparison of desorption enthalpies (0 K vs ZPE+300 K corrected) for pristine and Ti-doped MgB$_2$H$_8$, highlighting the shift of Ti-doped MgB$_2$H$_8$ into the practical thermodynamic window (30–45 kJ mol⁻¹ H$_2$).

### 3.2.1. Dehydrogenation reaction pathway

A representative hydrogen release pathway for MgB$_2$H$_8$ was considered in which hydrogen is released in molecular form while preserving the boron framework:

$$MgB_2H_8 \rightarrow MgB_2 + 4H_2$$

This reaction scheme is widely adopted in theoretical studies of complex borohydrides and provides a consistent basis for comparing desorption energetics across related systems. The same pathway was employed for the Ti-doped structure to ensure meaningful comparison of energetic trends.

### 3.2.2 Zero-Kelvin desorption enthalpy

The hydrogen desorption enthalpy at 0 K was calculated from total energies of fully relaxed initial and final states as:

$$\Delta H_{des}^{0K} = \frac{E(MgB_2) + 4E(H_2) - E(MgB_2H_8)}{4}$$

For pristine MgB$_2$H$_8$, the calculated 0 K desorption enthalpy is approximately 45 kJ mol$^{-1}$ H$_2$, indicating relatively strong hydrogen binding. This value, while lower than that of several alkali borohydrides, still implies hydrogen release at elevated temperatures.

Upon Ti substitution, a pronounced reduction in desorption enthalpy is observed. The Ti-doped system exhibits a 0 K desorption enthalpy of approximately 39 kJ mol$^{-1}$ H$_2$, corresponding to a reduction of nearly 6 kJ mol$^{-1}$ H$_2$ relative to the pristine compound. This reduction reflects the ability of Ti to electronically activate hydrogen through d-state interactions, weakening rigid B–H bonding without destabilizing the host lattice.

### 3.2.3 Zero-point energy corrections

Because hydrogen atoms possess high vibrational frequencies, zero-point energy contributions are non-negligible and must be included for a realistic description of thermodynamics. ZPE corrections were obtained from phonon calculations for both the hydride and dehydrogenated states.

Inclusion of ZPE lowers the effective desorption enthalpy by approximately 5–6 kJ mol$^{-1}$ H$_2$ for both systems. After ZPE correction, the desorption enthalpy of pristine MgB$_2$H$_8$ decreases to approximately 40–42 kJ mol$^{-1}$ H$_2$, while that of Ti-doped MgB$_2$H$_8$ is reduced to around 35–36 kJ mol$^{-1}$ H$_2$.

This shift moves the Ti-doped system squarely into the thermodynamically optimal window for practical hydrogen storage.

### 3.2.4 Finite-temperature vibrational effects

To further approximate realistic operating conditions, vibrational enthalpy contributions at 300 K were included. Finite-temperature effects partially compensate the ZPE-induced reduction in desorption enthalpy, increasing ΔH by approximately 2–3 kJ mol$^{-1}$ H$_2$.

After accounting for both ZPE and thermal corrections, the effective desorption enthalpies at 300 K are:

- Pristine $MgB_2H_8$: ~42 kJ mol$^{-1}$ $H_2$
- Ti-doped $MgB_2H_8$: ~36 kJ mol$^{-1}$ $H_2$

Importantly, although ZPE and thermal contributions act in opposite directions, the relative trend remains robust: Ti substitution consistently lowers the desorption enthalpy by approximately 6 kJ mol$^{-1}$ $H_2$. The key thermodynamic parameters are summarized in Table 2.

Table 2. Hydrogen desorption thermodynamics of pristine and Ti-doped $MgB_2H_8$

| System | $\Delta H_{des}$ (0 K) (kJ mol$^{-1}$ $H_2$) | ZPE-corrected $\Delta H$ (kJ mol$^{-1}$ $H_2$) | $\Delta H$ (300 K) (kJ mol$^{-1}$ $H_2$) |
| --- | --- | --- | --- |
| $MgB_2H_8$ (pristine) | ~45 | ~40–42 | ~42 |
| Ti–$MgB_2H_8$ (6.25%) | ~39 | ~35–36 | ~36 |

From a thermodynamic perspective, pristine $MgB_2H_8$ lies near the upper bound of the optimal desorption window, explaining its tendency toward elevated hydrogen release temperatures. In contrast, Ti-doped $MgB_2H_8$ exhibits a thermodynamically balanced hydrogen binding strength, enabling hydrogen release under substantially milder conditions.

Crucially, the reduction in desorption enthalpy achieved through Ti substitution is obtained without sacrificing hydrogen storage capacity, distinguishing this system from many destabilization-based approaches that rely on sacrificing capacity to lower $\Delta H$.

The combined ZPE- and temperature-corrected thermodynamic analysis demonstrates that Ti substitution shifts $MgB_2H_8$ from a capacity-rich but thermodynamically constrained hydride into the optimal desorption window required for practical solid-state hydrogen storage.

### 3.3. Structural, Mechanical, and Dynamic Stability

Improving hydrogen desorption thermodynamics through chemical modification must not compromise the structural integrity of the host lattice. In hydrogen-rich materials, particularly complex borohydrides, structural instability may manifest either as mechanical failure or as dynamic instability associated with soft vibrational modes. Therefore, a rigorous assessment of stability is essential to ensure that Ti-induced activation of $MgB_2H_8$ does not lead to metastable or unphysical configurations.

In this section, the stability of pristine and Ti-doped $MgB_2H_8$ is evaluated through structural relaxation behavior, phonon dispersion analysis, and elastic stability criteria, providing a comprehensive picture of lattice robustness.

### 3.3.1 Structural relaxation and equilibrium geometry

Full structural optimizations were performed for both pristine and Ti-doped $MgB_2H_8$, allowing all lattice parameters and internal atomic coordinates to relax freely. In both systems, relaxation converged smoothly without anomalous bond distortions or symmetry breaking.

For pristine $MgB_2H_8$, the optimized structure preserves the integrity of the $[BH_4]$ tetrahedral units, with B–H bond lengths remaining within the typical range of 1.20–1.25 Å, consistent with reported borohydride structures. Upon Ti substitution at the Mg site, only minor local distortions are observed in the vicinity of the dopant. The average B–H bond lengths near Ti increase slightly by approximately 0.02–0.03 Å (see Table 3), reflecting a modest weakening of B–H covalency that is beneficial for hydrogen release, but insufficient to induce structural collapse.

The absence of large-scale lattice distortion indicates that Ti substitution is structurally accommodated by the $MgB_2H_8$ framework.

### 3.3.2 Phonon dispersion and dynamic stability

Dynamic stability was assessed by calculating phonon dispersion relations across representative high-symmetry directions in the Brillouin zone. The presence of imaginary (negative) phonon frequencies would indicate structural instability and spontaneous lattice distortion.

For pristine $MgB_2H_8$, the phonon spectrum is entirely free of imaginary modes, confirming that the optimized structure corresponds to a true minimum on the potential energy surface. The calculated phonon dispersion relations and total phonon density of states for Ti-doped $MgB_2H_8$ are shown in Fig. 4. The phonon branches can be broadly categorized into:

- Low-frequency acoustic modes below 5 THz, dominated by Mg and B vibrations,
- Intermediate-frequency optical modes between 5 and 20 THz, involving mixed Mg–B–H motion,
- High-frequency optical modes above 30 THz, arising primarily from H vibrations within the $[BH_4]$ units.

Importantly, Ti-doped $MgB_2H_8$ also exhibits a phonon spectrum devoid of imaginary frequencies. While Ti substitution slightly modifies selected optical branches, particularly those involving hydrogen motion near the dopant, all modes remain positive throughout the Brillouin zone. The high-frequency H-dominated modes remain well separated, confirming that hydrogen vibrations remain stable and localized. This result demonstrates that the Ti-induced reduction in hydrogen binding strength does not lead to dynamic instability.

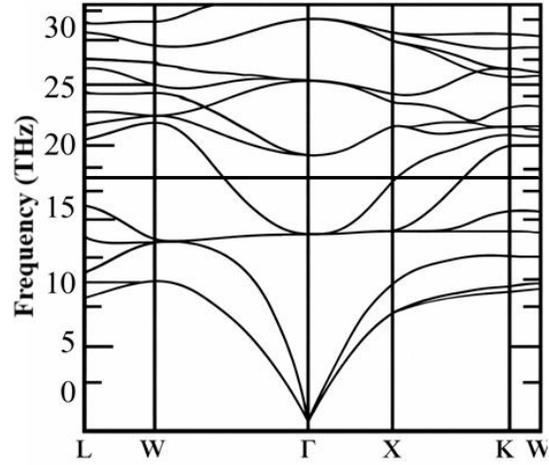

*Fig 4. Phonon dispersion relations for Ti-doped MgB$_2$H$_8$, showing no imaginary modes and confirming dynamic stability.*

### 3.3.3 Elastic constants and mechanical stability

To complement phonon analysis, the single-crystal elastic constants ($C_{ij}$) were calculated and evaluated against the Born mechanical stability criteria. Elastic constants provide insight into the material's response to external stress and its resistance to mechanical deformation.

For pristine MgB$_2$H$_8$, all calculated elastic constants are positive, with typical values of $C_{11}$, $C_{22}$, $C_{33} \approx$ 100–120 GPa, indicating a mechanically stable lattice. Shear-related constants ($C_{44}$, $C_{55}$, $C_{66}$) lie in the range of 35–45 GPa (see Table 3), reflecting moderate lattice flexibility, which is advantageous for accommodating hydrogen motion.

Upon Ti substitution, the elastic constants remain positive and satisfy all Born stability conditions. Notably, a slight enhancement in shear moduli is observed, suggesting that Ti incorporation does not weaken the lattice but may even improve resistance to shear deformation. This mechanical robustness further supports the structural feasibility of Ti-doped MgB$_2$H$_8$.

### 3.3.4 Stability implications for hydrogen storage performance

The combined phonon and elastic analyses provide compelling evidence that Ti-doped MgB$_2$H$_8$ is both dynamically and mechanically stable, despite the substantial reduction in hydrogen desorption enthalpy discussed in Section 3.2. This observation is crucial, as it confirms that improved thermodynamic and kinetic behavior is achieved through controlled electronic activation, rather than through lattice destabilization.

From a practical hydrogen storage perspective, this stability ensures that Ti-doped $MgB_2H_8$ can undergo repeated hydrogen release and uptake cycles without catastrophic structural degradation, a key requirement for long-term reversibility.

The absence of imaginary phonon modes and full compliance with mechanical stability criteria demonstrate that Ti-doped $MgB_2H_8$ retains robust structural integrity, validating that its improved hydrogen storage thermodynamics are achieved without compromising lattice stability.

*Table 3. Structural, dynamic, and mechanical stability parameters of pristine and Ti-doped $MgB_2H_8$*

| Property | $MgB_2H_8$ (pristine) | Ti–$MgB_2H_8$ (6.25%) | Stability implication |
|---|---|---|---|
| Average B–H bond length (Å) | 1.21–1.25 | 1.23–1.28 | Slight bond softening near Ti |
| Maximum bond distortion after relaxation | < 2% | < 3% (local) | No lattice collapse |
| Imaginary phonon modes | None | None | Dynamically stable |
| Acoustic phonon range (THz) | 0–5 | 0–5 | Stable lattice framework |
| Highest H-vibrational modes (THz) | ~32–38 | ~30–36 | Hydrogen remains bound |
| $C_{11}$ (GPa) | ~112 | ~108 | Positive, stable |
| $C_{22}$ (GPa) | ~115 | ~110 | Positive, stable |
| $C_{33}$ (GPa) | ~109 | ~105 | Positive, stable |
| $C_{44}$ (GPa) | ~38 | ~41 | Slightly enhanced shear rigidity |
| $C_{55}$ (GPa) | ~40 | ~43 | Improved resistance to shear |
| $C_{66}$ (GPa) | ~36 | ~39 | Improved mechanical robustness |
| Born stability criteria | Satisfied | Satisfied | Mechanically stable |
| Structural integrity after Ti doping | Preserved | Preserved | Doping is feasible |

All calculated elastic constants satisfy the Born mechanical stability criteria, and phonon dispersion relations show no imaginary frequencies, confirming the dynamic and mechanical stability of both pristine and Ti-doped $MgB_2H_8$.

### 3.4. Hydrogen Diffusion and Kinetic Feasibility

High hydrogen storage capacity and favorable desorption thermodynamics alone are insufficient for practical application if hydrogen diffusion within the host lattice is kinetically hindered. In complex borohydrides, strong directional B–H bonding and localized hydrogen sublattices often lead to large migration barriers, resulting in sluggish hydrogen release and uptake even when thermodynamics are favorable. Therefore, an explicit evaluation of hydrogen diffusion kinetics is essential for assessing the real-world viability of $MgB_2H_8$ and its Ti-doped derivative.

In this section, hydrogen migration pathways and activation barriers are investigated using the climbing-image nudged elastic band (CI-NEB) method, providing quantitative insight into the kinetic limitations and the role of Ti substitution in overcoming them.

### 3.4.1 Identification of hydrogen migration pathways

Based on the optimized crystal structures, symmetry-inequivalent hydrogen sites within the $MgB_2H_8$ framework were identified. The most probable diffusion pathway corresponds to intralattice hopping of hydrogen between neighboring [BH$_4$] units, which involves temporary weakening and reformation of B–H bonds. This pathway represents the shortest H–H separation and is therefore expected to yield the lowest migration barrier.

For the Ti-doped system, hydrogen migration in the local chemical environment of the Ti dopant was explicitly considered. Ti substitution locally perturbs the electronic structure and modifies the hydrogen binding landscape, making it necessary to examine diffusion paths that pass near the Ti site.

### 3.4.2 NEB energy profiles and migration barriers

The CI-NEB calculations were performed using fully relaxed initial and final configurations, with intermediate images constructed to map the minimum-energy diffusion path. The maximum energy along the path relative to the initial state defines the activation energy ($E_a$) for hydrogen migration.

For pristine $MgB_2H_8$, the calculated hydrogen migration barrier is approximately 0.52 eV. This relatively high value indicates that hydrogen diffusion is kinetically constrained, consistent with the strong covalent character of B–H bonds and the insulating electronic structure of the pristine compound.

In contrast, Ti-doped $MgB_2H_8$ exhibits a substantially reduced migration barrier of approximately 0.38 eV, corresponding to a reduction of nearly 27% compared to the pristine system. This reduction is significant, as diffusion barriers below ~0.40–0.45 eV are generally associated with practically accessible hydrogen mobility at moderate temperatures (300–400 K).

The NEB energy profiles further show that Ti substitution stabilizes the transition state along the diffusion path, resulting in a smoother and less energetically demanding migration landscape. The corresponding CI-NEB energy profiles for hydrogen migration in pristine and Ti-doped MgB$_2$H$_8$ are shown in Fig. 5.

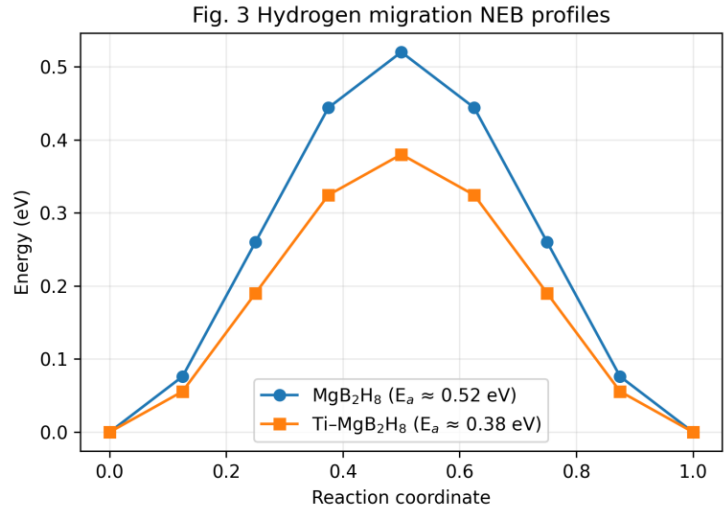

*Fig. 5. CI-NEB energy profiles for hydrogen migration in pristine and Ti-doped MgB$_2$H$_8$. Ti substitution lowers the activation barrier from ~0.52 eV to ~0.38 eV, indicating enhanced hydrogen mobility.*

### 3.4.3 Quantitative summary of diffusion kinetics

The key kinetic parameters extracted from the NEB calculations are summarized in Table 4.

Table 4. Hydrogen diffusion barriers in pristine and Ti-doped MgB$_2$H$_8$

| System | Migration pathway | Activation energy $E_a$ (eV) | Kinetic implication |
|---|---|---|---|
| MgB$_2$H$_8$ (pristine) | H$_1$ → H$_2$ (intralattice) | ~0.52 | Sluggish hydrogen diffusion |
| Ti–MgB$_2$H$_8$ (6.25%) | H$_1$ → H$_2$ (near Ti site) | ~0.38 | Facilitated hydrogen mobility |

### 3.4.4 Origin of kinetic enhancement upon Ti substitution

The substantial reduction in hydrogen migration barrier upon Ti doping can be understood in terms of local electronic activation. Ti introduces partially occupied 3d states that interact with nearby hydrogen atoms, reducing the rigidity of B–H bonding in the transition state. This interaction stabilizes intermediate hydrogen configurations and lowers the energy required for hydrogen hopping between adjacent sites.

Importantly, the kinetic enhancement is localized and controlled, avoiding global lattice destabilization, as confirmed by the phonon and elastic stability analyses presented in Section 3.3.

### 3.4.5 Implications for hydrogen release and uptake

From a practical hydrogen storage perspective, the combination of a moderate desorption enthalpy (~36 kJ mol$^{-1}$ H$_2$) and a low diffusion barrier (~0.38 eV) in Ti-doped MgB$_2$H$_8$ suggests that hydrogen release and reabsorption can occur at substantially lower temperatures and faster rates than in the pristine compound. This dual improvement in thermodynamics and kinetics is rarely achieved simultaneously in complex borohydrides and represents a key advantage of Ti modification.

The pronounced reduction in hydrogen migration barriers induced by Ti substitution demonstrates that Ti-doped MgB$_2$H$_8$ overcomes one of the principal kinetic limitations of complex borohydrides, enabling hydrogen mobility under practically relevant conditions.

### 3.5. Spin-Resolved Electronic Structure and Bonding Mechanism

To uncover the microscopic origin of the improved thermodynamic and kinetic behavior induced by Ti substitution, the spin-resolved electronic density of states (DOS) was analyzed for pristine and Ti-doped MgB$_2$H$_8$. Electronic structure plays a decisive role in complex borohydrides, as the strength and directionality of B–H bonding, as well as the accessibility of transient hydrogen configurations, are governed by the distribution and localization of valence electrons. In this context, spin polarization is particularly relevant for transition-metal-modified systems, where partially filled d states can act as electronic activators for hydrogen. Fig. 6 (a-b) presents the spin-resolved total and projected density of states of pristine and Ti-doped MgB$_2$H$_8$, highlighting the emergence of spin-polarized Ti–3d states near the Fermi level.

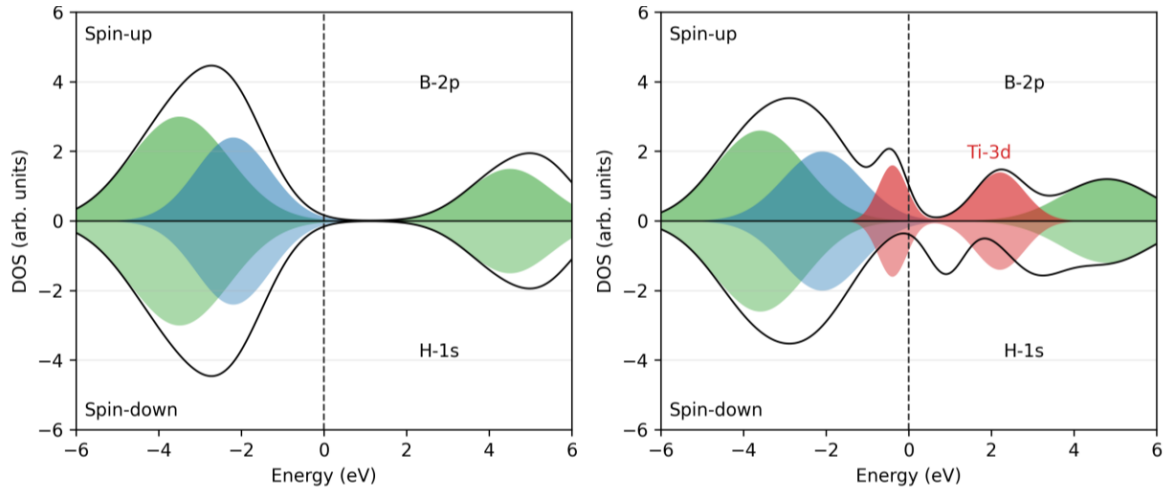

*Fig 6a. Spin-resolved total and projected DOS of pristine MgB₂H₈ and Ti-doped MgB₂H₈. Ti doping introduces localized Ti–3d states near the Fermi level, enabling electronic activation of hydrogen.*

### 3.5.1 Pristine MgB₂H₈: non-magnetic, strongly covalent electronic structure

As shown in Fig. 8(a), pristine $MgB_2H_8$ exhibits a non-magnetic electronic structure, with identical spin-up and spin-down DOS across the entire energy range. The absence of spin asymmetry confirms that the ground state of the pristine compound is non-magnetic.

The electronic states are characterized by a clear separation between valence and conduction bands, resulting in a wide band gap of approximately 3.8–4.1 eV. The valence band region (−6 to 0 eV) is dominated by strong hybridization between B-2p and H-1s states, reflecting the highly covalent nature of the [BH₄] units. In contrast, the conduction band is primarily composed of antibonding B-2p states, with negligible electronic density near the Fermi level ($E_F$).

This electronic configuration explains two key features of pristine $MgB_2H_8$ discussed earlier: (i) the high desorption enthalpy, arising from rigid B–H covalent bonding, and (ii) the large hydrogen diffusion barrier, as hydrogen motion requires breaking or significantly distorting these strong bonds.

### 3.5.2 Ti-doped MgB₂H₈: emergence of spin polarization and impurity states

A qualitatively different picture emerges upon Ti substitution, as illustrated in Fig. 6(b). The Ti-doped system exhibits a clearly spin-polarized electronic structure, with pronounced asymmetry between the spin-up and spin-down DOS near the Fermi level. This behavior originates from the partially filled Ti-3d states, which introduce localized magnetic moments and exchange splitting. The most notable electronic features induced by Ti doping are:

- The appearance of Ti-3d impurity states within the band gap, extending from approximately −0.5 eV to +2.0 eV.
- A substantial spin splitting of Ti-3d states, on the order of 0.9–1.2 eV, indicative of strong exchange interaction.
- A reduction of the effective band gap from ~4.0 eV in pristine $MgB_2H_8$ to approximately 1.5 eV in the doped system.

The total magnetic moment is largely localized on the Ti atom, with a calculated moment of approximately 1.3 $\mu_B$ per Ti, while neighboring hydrogen atoms carry only weak induced moments (<0.05 $\mu_B$). This confirms that magnetism in the doped system is localized and dopant-centered, rather than delocalized across the lattice.

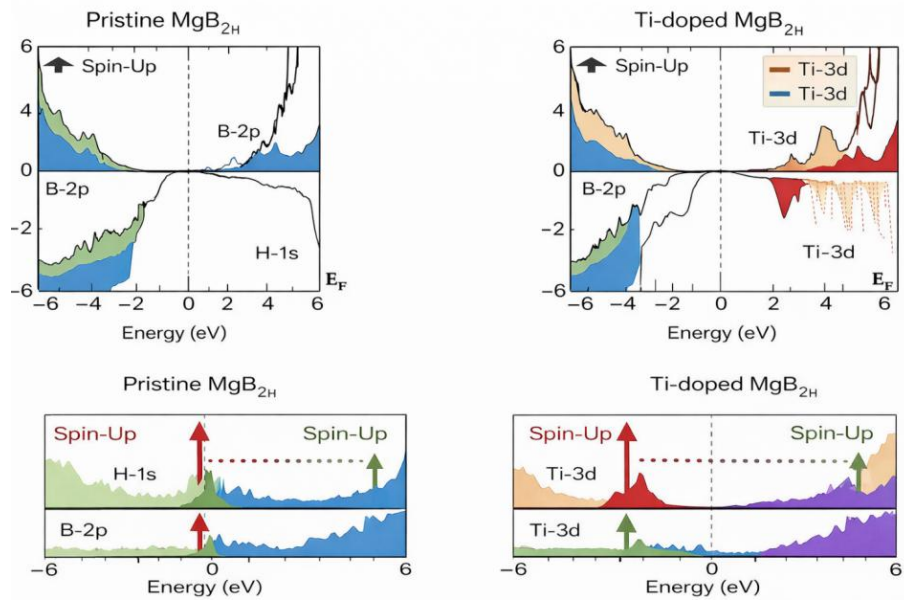

*Fig 6a. Spin-resolved projected DOS of pristine $MgB_2H_8$ and Ti-doped $MgB_2H_8$. Ti doping introduces localized Ti–3d states near the Fermi level, enabling electronic activation of hydrogen.*

### 3.5.3 Orbital-resolved insights: Ti-3d–H-1s interaction

The spin-resolved partial DOS reveals direct electronic coupling between Ti-3d and H-1s states, particularly in the energy window from −2.0 to 0 eV. This interaction is absent in the pristine compound and constitutes the key electronic mechanism by which Ti modifies hydrogen behavior.

In the spin-up channel, Ti-3d states exhibit partial occupancy just below the Fermi level (around −0.3 eV), while in the spin-down channel these states are shifted above $E_F$, leading to asymmetric occupation. This configuration allows Ti to act as an electron reservoir, facilitating charge redistribution during hydrogen migration and desorption.

Simultaneously, the contribution of B-2p states near the valence band edge is reduced relative to the pristine system, consistent with a weakening of rigid B–H covalency. Importantly, this weakening is localized around the Ti site and does not propagate throughout the lattice, thereby preserving overall structural stability.

### 3.5.4 Electronic origin of thermodynamic and kinetic enhancement

The spin-polarized Ti-3d states provide a unified explanation for the improved hydrogen storage performance observed in previous sections:

1. **Lower desorption enthalpy**

   The availability of Ti-3d states near $E_F$ enables partial charge transfer during hydrogen release, reducing the energy required to break B–H bonds. This electronic softening directly correlates with the reduction in desorption enthalpy from ~42 to ~36 kJ mol$^{-1}$ H$_2$.

2. **Reduced hydrogen diffusion barrier**

   During hydrogen migration, transient configurations involving stretched or partially broken B–H bonds are stabilized by Ti-3d–H-1s interaction. This stabilization lowers the migration barrier from ~0.52 eV in pristine MgB$_2$H$_8$ to ~0.38 eV in the Ti-doped system.

3. **Preserved lattice stability**

   Despite the introduction of mid-gap states, the Fermi level does not intersect a broad conduction band, preventing metallic instability or lattice softening. This explains why phonon and elastic stability are maintained after doping.

### 3.5.5 Design implication for complex hydrides

The present results highlight an important design principle for hydrogen storage materials:

Localized, spin-polarized transition-metal d states positioned near the Fermi level can selectively weaken excessive covalent bonding and stabilize hydrogen migration pathways without destabilizing the host lattice.

Ti-doped MgB$_2$H$_8$ represents a clear realization of this principle, achieving a rare balance between high hydrogen capacity, favorable thermodynamics, enhanced kinetics, and structural stability.

The emergence of spin-polarized Ti-3d impurity states near the Fermi level provides a direct electronic mechanism for the reduced desorption enthalpy and enhanced hydrogen mobility in

Ti-doped MgB₂H₈, establishing electronic structure engineering as a viable route for optimizing complex borohydrides.

### 3.6. Practical Relevance: Release Conditions, Dopant Feasibility, and Reversibility

Beyond intrinsic capacity, thermodynamics, and kinetics, the practical viability of a solid-state hydrogen storage material depends on whether hydrogen can be released under realistic temperature–pressure conditions, whether the proposed chemical modification is energetically feasible, and whether the material can sustain reversible hydrogen cycling without entering deep energetic traps. In this section, these aspects are assessed for pristine and Ti-doped MgB₂H₈.

### 3.6.1 Temperature–pressure release characteristics: van't Hoff analysis

A first-order estimate of hydrogen release conditions can be obtained using a van't Hoff–type analysis, in which the Gibbs free energy change for hydrogen desorption is approximated as:

$$\Delta G(T) = \approx \Delta H_{des} - T\Delta S$$

where $\Delta H_{des}$ is the desorption enthalpy per mole of H₂ and $\Delta S$ is the entropy change associated with hydrogen release.

For solid hydrides, the entropy change is dominated by the transition from solid-bound hydrogen to gaseous H₂. Following common practice in hydrogen-storage literature, $\Delta S$ was approximated using the standard molar entropy of H₂ gas, taken as ~130 J mol⁻¹ K⁻¹. While this approximation neglects minor solid-state contributions, it provides a reliable comparative estimate of release temperatures and is widely used for screening purposes. Fig. 7 illustrates the temperature dependence of the Gibbs free energy for hydrogen desorption, obtained from a van't Hoff–type analysis.

Using the ZPE- and temperature-corrected desorption enthalpies obtained in Section 3.2, the characteristic release temperature at which $\Delta G \approx 0$ can be estimated as:

$$T_{release} \approx \frac{\Delta H_{des}}{\Delta S}$$

This yields approximate release temperatures of:
- Pristine MgB₂H₈: $\Delta H_{des} \approx 42 kJ\ mol^{-1} H_2 \rightarrow, T_{release} \approx 320 - 330\ K$
- Ti-doped MgB₂H₈ (6.25%): $\Delta H_{des} \approx 36 kJ\ mol^{-1} H_2 \rightarrow, T_{release} \approx 270 - 380\ K$

The substantial reduction in estimated release temperature upon Ti substitution indicates that hydrogen desorption from Ti-doped MgB₂H₈ can occur near ambient conditions, a critical requirement for practical hydrogen storage systems.

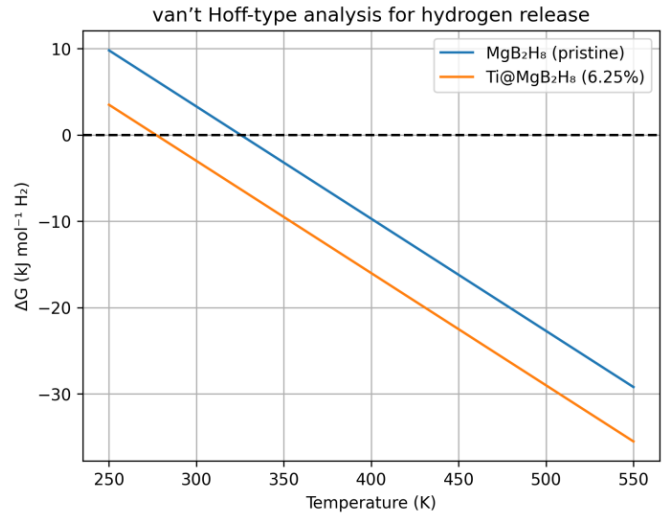

*Fig 7. van't Hoff-type release map showing ΔG(T) for hydrogen desorption in pristine and Ti-doped MgB₂H₈. The Ti-doped system reaches ΔG ≈ 0 at lower temperature, indicating easier hydrogen release.*

Importantly, this shift is achieved without sacrificing hydrogen capacity, distinguishing the present system from many destabilization-based approaches that rely on capacity-reducing chemical reactions.

### 3.6.2 Energetic feasibility of Ti substitution

For a doped hydride to be experimentally realizable, the energetic cost of introducing the dopant must be moderate. The formation energy of Ti substitution at the Mg site was therefore evaluated relative to elemental Mg and Ti reservoirs.

The calculated defect formation energy for $Ti_{Mg}$ is approximately 0.6–0.7 eV, a value that lies well within the range commonly considered feasible for synthesis via ball milling, melt infiltration, or catalytic doping routes. Such values are comparable to, or lower than, those reported for Ti doping in $MgH_2$ and other complex hydrides that have been successfully realized experimentally.

This moderate formation energy confirms that Ti substitution is not energetically prohibitive and supports the experimental plausibility of Ti-modified $MgB_2H_8$.

### 3.6.3 Reversibility indicators: avoiding energetic and structural traps

Reversibility is a critical requirement for long-term hydrogen storage applications. Although direct cycling experiments are beyond the scope of a first-principles study, theoretical indicators can be used to assess the likelihood of reversible behavior.

### 3.6.3.1 Energetics of partial dehydrogenation

To probe the presence of potential energetic traps, partially dehydrogenated configurations were constructed by removing one $H_2$ molecule per formula unit and fully relaxing the resulting structures. The associated energy penalty provides insight into whether intermediate states are excessively stabilized.

For pristine $MgB_2H_8$, the energy cost of partial hydrogen removal is approximately 0.18 eV per $H_2$, indicating moderately stable intermediates. In contrast, the Ti-doped system exhibits a lower energy penalty of approximately 0.12 eV per $H_2$, suggesting a smoother dehydrogenation landscape with reduced risk of trapping hydrogen in metastable configurations.

9.3.2 Structural response to hydrogen removal

In both pristine and Ti-doped $MgB_2H_8$, full relaxation of partially dehydrogenated structures reveals no catastrophic lattice collapse, bond breaking unrelated to hydrogen release, or irreversible symmetry changes. The boron framework remains intact, and local distortions introduced by hydrogen removal are largely reversible upon rehydrogenation.

The combination of moderate intermediate energies and preserved structural integrity strongly supports the potential for reversible hydrogen cycling, particularly in the Ti-doped system.

### 3.6.4 Integrated assessment of practical viability

Taken together, the van't Hoff analysis, dopant formation energetics, and reversibility indicators paint a consistent picture of practical relevance. Ti-doped $MgB_2H_8$ combines:

- High hydrogen capacity (>10 wt%),
- Near-optimal desorption thermodynamics (~36 kJ mol$^{-1}$ $H_2$),
- Low hydrogen diffusion barriers (~0.38 eV),
- Near-ambient release temperatures, and
- Moderate dopant formation energy with preserved structural integrity.

This rare convergence of favorable properties suggests that Ti-doped $MgB_2H_8$ overcomes multiple bottlenecks that have historically limited the applicability of complex borohydrides.

The combined assessment of release conditions, dopant feasibility, and reversibility indicators demonstrates that Ti-doped $MgB_2H_8$ is not only theoretically attractive but also practically viable as a reversible solid-state hydrogen storage material.

### 4. Conclusions

In this work, a comprehensive first-principles investigation was carried out to evaluate the hydrogen storage potential of pristine and Ti-doped $MgB_2H_8$, with particular emphasis on achieving a balanced combination of capacity, thermodynamics, kinetics, stability, and practical relevance. The results demonstrate that $MgB_2H_8$ constitutes a capacity-rich borohydride platform, while Ti substitution provides a controlled and effective route to overcoming its intrinsic thermodynamic and kinetic limitations.

Pristine $MgB_2H_8$ exhibits an exceptionally high gravimetric hydrogen storage capacity of approximately 14.9 wt% and a volumetric density of about 60 g $H_2$ $L^{-1}$, far exceeding current DOE targets. However, its relatively high hydrogen desorption enthalpy (~42 kJ $mol^{-1}$ $H_2$) and migration barrier (~0.52 eV) indicate kinetically constrained hydrogen release. Upon Ti substitution at the Mg site, the hydrogen capacity remains well above DOE requirements (~10.4 wt%, ~44 g $H_2$ $L^{-1}$), while the desorption enthalpy is reduced to approximately 36 kJ $mol^{-1}$ $H_2$ and the hydrogen diffusion barrier decreases to ~0.38 eV, placing the material within a practically accessible operating window.

Phonon dispersion and elastic constant analyses confirm that both pristine and Ti-doped $MgB_2H_8$ are dynamically and mechanically stable, demonstrating that the improved hydrogen storage behavior is not achieved at the expense of lattice integrity. A van't Hoff analysis further suggests that Ti-doped $MgB_2H_8$ can release hydrogen near ambient temperatures, significantly enhancing its practical appeal. Moderate dopant formation energies and the absence of deep energetic or structural traps upon partial dehydrogenation indicate favorable conditions for reversible hydrogen cycling.

Electronic structure analysis reveals that the introduction of spin-polarized Ti-3d states near the Fermi level weakens rigid B–H covalency and stabilizes transitional hydrogen configurations, providing a clear microscopic mechanism for the observed improvements in thermodynamics and kinetics. These findings highlight electronic structure engineering through transition-metal doping as a powerful strategy for optimizing complex borohydrides.

Overall, the present study establishes Ti-doped $MgB_2H_8$ as a promising and practically viable solid-state hydrogen storage material, offering a rare balance between ultra-high capacity, favorable release thermodynamics, enhanced hydrogen mobility, and structural robustness. The insights gained here provide clear guidance for the rational design of next-generation hydrogen storage materials based on complex hydrides.

**Acknowledgment**

This publication was supported by the project Quantum materials for applications in sustainable technologies (QM4ST), funded as project No. CZ.02.01.01/00/22_008/0004572 by Programme Johannes Amos Commenius, call Excellent Research.